\documentclass[aps,prb,twocolumn,groupedaddress,showpacs]{revtex4-1}

\usepackage{graphicx}
\usepackage{hyperref}
\usepackage{hypcap}

\begin{document}

%\preprint{}

\title{Scanning tunneling spectroscopy of a dilute two-dimensional electron system exhibiting Rashba spin splitting}


\author{S. Becker}
\email[]{stefan.becker@physik.rwth-aachen.de}
%\thanks{}
\author{M. Liebmann}
\author{T. Mashoff}
\author{M. Pratzer}
\author{M. Morgenstern}
\affiliation{II. Physikalisches Institut B and JARA-FIT, RWTH Aachen University, 52074 Aachen, Germany}
\date{\today}

\begin{abstract}
Using scanning tunneling spectroscopy (STS) at $5\,\text{K}$ in $B$-fields up to $7\,\text{T}$, we investigate the local density of states of a two-dimensional electron system (2DES) created by Cs adsorption on $p$-type InSb(110). The 2DES, which in contrast to previous STS studies exhibits a 2D Fermi level, shows standing waves at $B=0\,\text{T}$ with corrugations decreasing with energy and with wave numbers in accordance with theory. In magnetic field percolating drift states are observed within the disorder broadened Landau levels. Due to the large electric field perpendicular to the surface, a beating pattern of the Landau levels is found and explained quantitatively by Rashba spin splitting within the lowest 2DES subband. The Rashba splitting does not contribute significantly to the standing wave patterns in accordance with theory.
\end{abstract}
%\keywords{}

\maketitle

\section{Introduction}
The spin-orbit coupling in semiconductors provides the opportunity to manipulate spins by electric fields,\cite{datta:665} which might become a central handle in spintronics.\cite{Wolf,Fabian} The corresponding spin-splitting is called the Rashba effect.\cite{Rashba,BychkovRashba} It has been probed in III-V-semiconductors, e.g., by the beating pattern of Shubnikov-de~Haas oscillations\cite{Nitta, Grundler} or by the analysis of weak antilocalization.\cite{Miller,Nitta2} The Rashba effect has also been probed on metal surfaces by angular resolved photoelectron spectroscopy\cite{LaShell,Hoesch,Ast,Dedkov} as well as by scanning tunneling spectroscopy (STS).\cite{Ast2,Pascual} In the latter case, either the enhancement of the density of states close to the onset of the two-dimensional surface band\cite{Ast2} or the suppression of quasiparticle interference caused by a missing spin-Umklapp scattering\cite{Pascual} has been used to deduce the Rashba effect indirectly. Theoretical considerations reveal that the Rashba splitting is not directly visible in quasiparticle interference patterns probed by STS, if only single scattering is considered.\cite{Pettersen} Subtle changes of the quasiparticle interference appear, if multiple scattering becomes relevant.\cite{Walls}

So far, Rashba spin splitting within semiconductors has not been probed in real space. Here, we use the surface doping effect, which leads to a two-dimensional electron system (2DES) directly at the surface of a low-gap III-V semiconductor to be probed by STS.\cite{Aristov, FeInAs, NbInAs, PhysRevB.63.155315} These 2DESs exhibit a quantum Hall effect down to $B=2\,\text{T}$ (filling factor 7) and a reasonable mobility of $\mu=6000\,\text{cm}^2/(\text{V\,s})$.\cite{Masutomi,Tsuji} The low gap provides, in addition, a relatively large Rashba splitting\cite{PhysRevB.61.15588} as well as a strong Landau\cite{Morgenstern3} and spin splitting.\cite{Morgenstern4} This 2DES has been probed previously by STS revealing transitions from strong to weak localization,\cite{PhysRevB.68.041402} a spatially continuous wave pattern caused by multiple scattering,\cite{PhysRevLett.89.136806} drift states in magnetic field\cite{Morgenstern2} and the local density of states across quantum Hall transitions.\cite{hashimoto:256802}

Within this paper, we describe STS measurements of a 2DES, which is induced by a minute amount of Cs ($1.5\,\%$ of a monolayer) on the strongly $p$-doped InSb(110) surface ($N_{\text{A}} \simeq 10^{24}\,\text{m}^{-2}$). The strong doping results in a strong electric field ($\ge 10^7\,\text{V}/\text{m}$) within the 2DES and correspondingly leads to a large Rashba coefficient ($\alpha \simeq 10^{-10} \,\text{eV}\,\text{m}$). We demonstrate that this 2DES still exhibits spatially continuous wave patterns at $B=0\,\text{T}$ with preferential wave vectors in agreement with $\textbf{k}\cdot \textbf{p}$-theory as well as strong corrugation ($50\,\%$ at $E_{\text{F}}$), very similar to the 2DES prepared on $n$-type InAs(110).\cite{PhysRevLett.89.136806} Also Landau levels and drift states are observed by STS in a $B$-field.\cite{Morgenstern2} The disorder given by the acceptors prohibits the observation of spin splitting within the spatially averaged density of states, but spin splitting is observed in the local density of states ($LDOS(x,y)$). Importantly, the Rashba effect leads to a pronounced beating of the Landau level intensity measured by STS, which is quantitatively reproduced by calculations using $\alpha = 7 \times 10^{-11}\,\text{eV}\,\text{m}$. This experimental value is very close to the expected value of $\alpha = 9\text{--}11 \times 10^{-11}\,\text{eV}\,\text{m}$ deduced from the known surface band bending. Thus, we demonstrate that the Rashba parameter can be determined down to the nm scale.

\section{Experiment}
Our home-built scanning tunneling microscope (STM) operates within an ultra-high vacuum (UHV) insert of a helium-4 bath cryostat at a base temperature of $5\,\text{Kelvin}$ and with a magnetic field up to $7\,\text{Tesla}$ perpendicular to the sample surface.\cite{mashoff:053702} For tip exchange and sample transfer, the STM can be lifted out of the cryostat into a standard UHV chamber without breaking the vacuum. A cesium dispenser (SAES Getters) mounted into this UHV chamber can dispense controlled amounts of cesium onto samples in the cold microscope. The STM tip was etched outside of the vacuum system from a tungsten wire and prepared within the STM by field emission and consecutive voltage pulses on a W(110) crystal. As a sample we used a Ge-doped InSb crystal with a hole concentration of $1\text{--}2 \times 10^{24}\,\text{m}^{-3}$ as determined by Hall measurements. In order to prepare the 2DES, the crystal was glued to a molybdenum sample holder with electrically conductive adhesive pointing with the (110) surface upwards. It was cleaved at room temperature at a background pressure of $p=2 \times 10^{-10}\,\text{mbar}$ resulting in a clean (110) surface. The sample was then transferred into the pre-cooled STM ($T < 60\,\text{K}$) and cesium was evaporated onto the crystal surface. Without cooling, the adsorbed cesium atoms would quickly form chain structures at room temperature\cite{whitman:770} breaking the homogeneous distribution. After transferring the microscope into the cryostat the sample could be probed for weeks without any noticeable change in adsorbate density. The cesium atoms appear as white dots in the STM image of Fig.~\ref{Wellen}(a), which shows a homogeneous Cs distribution amounting to about 1.5\,\% adsorbates per surface unit cell. STS data ($dI/dV$) were acquired by lock-in-technique applying a modulation voltage with amplitude $\sqrt{2} \times V_{\text{mod}}$ to the sample. Prior to measurement, the tip-surface distance is stabilized at voltage $V_{\text{stab}}$ and current $I_{\text{stab}}$. Then the feedback is turned off and the voltage is ramped to the measurement voltage $V_{\text{s}}$.

\section{Results and Discussion}
It is well known that adsorbates like alkali metals can induce band bending at semiconductor surfaces already at low adsorbate densities.\cite{Aristov} A strong downwards band bending on $p$-type semiconductors induces a so-called inversion-layer, which confines a 2DES directly below the surface.\cite{PhysRevB.61.15588} In contrast to 2DESs buried deeply in heterostructures,\cite{RevModPhys.54.437} this 2DES is accessible by STS, a technique, which measures the $LDOS(x,y)$ by the differential conductivity $dI/dV(x,y)$.\cite{SRL}

The strength of the adsorbate induced band bending mainly depends on the materials combination and the adsorbate coverage.\cite{Aristov, PhysRevB.63.155315, FeInAs, NbInAs} Cs on $n$-InSb(110) induces a maximum band bending of $V_{\text{bb}} = 290\,\text{meV}$ as determined by photoelectron spectroscopy,\cite{PhysRevB.63.155315} where $V_{\text{bb}}$ describes the energy difference between the Fermi level $E_{\text{F}}$ and the onset of the conduction band directly at the surface $E_{\text{CBM}}$. Therefore, the maximum band bending for Cs on $p$-InSb(110) is $V_{\text{bb}}+E_{\text{gap}}$ with the energy gap $E_{\text{gap}}=0.235\,\text{eV}$\cite{Vurgaftman} of InSb. The actual shape of the confining potential and the resulting subband energies of the 2DES further depend on the dopant concentration of the semiconductor. The steepness of the band bending increases with the acceptor concentration $N_{\text{A}}$. A large $N_{\text{A}}$ results in a steep confining potential and, thus, leads to large subband energies $E_i$, $i=0,1,2,\ldots$ measured with respect to $E_{\text{CBM}}$.  Consequently, one gets a low electron concentration within the inversion-layer $N_{\text{s}}$.

Assuming the subband energy levels $E_i$ to be known, the electron concentration in the inversion-layer $N_{\text{s}}$ is given by
\begin{eqnarray}
N_{\text{s}} &=& \sum_{i}\int^{E_{\text{F}}}_{E_{i}}D(E) dE,\label{eq:Ns}\\
D(E)&=&\frac{m^{*}(E)}{\pi\hbar^{2}},
\label{eq:DE}
\end{eqnarray}
where  $D(E)$ denotes the two-dimensional density of states including spin degeneracy and $m^{*}(E)$ is the effective electron mass within the conduction band, which, in general, is energy dependent due to non-parabolicity. An estimate for the so-called momentum effective mass or density of states effective mass for a 2DES defined as\cite{Zawadzki1974,Winkler.PhysRevB.48.8918}
\begin{eqnarray}
m^{*}=\hbar^2k\left(\frac{dE}{dk}\right)^{-1}
\label{eq:MomEffMass}
\end{eqnarray}
can be given within a triangular potential well approximation\cite{RevModPhys.54.437} using a $\textbf{k}\cdot\textbf{p}$ perturbation approach:\cite{KaneInSb,PhysRevB.35.2460}
\begin{eqnarray}
m^{*}(E) &=&	 m^{*}_{0} \left( 1 + 2 \frac{\frac{1}{3} E_{i} + E_{\parallel}}{E_{\text{gap}}}  \right).
\label{eq:Merkt}
\end{eqnarray}
Here, $E$ splits up into the in-plane energy $E_{\parallel}=E-E_{i}$ and the energy offset of the subband $E_{i}$ with respect to $E_{\text{CBM}}$, $m^{*}_{0}$ denotes the bulk effective electron mass at $E_{\text{CBM}}$ and $E_{\text{gap}}$ is the gap energy.\cite{PhysRevB.35.2460} Strictly speaking, this approach is limited to energies small compared to the gap energy. But although this is not the case for the 2DES induced by Cs on InSb, Eq.~(\ref{eq:Merkt}) can still be used to get a reasonable estimate of the effective mass.

\begin{figure}\capstart
\includegraphics{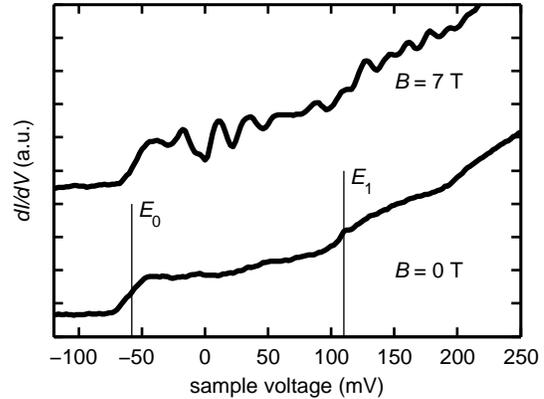}
\caption{Spatially averaged $dI/dV$ curves measured at $B = 0\,\text{T}$ and $B=7\,\text{T}$ as indicated, $V_{\text{stab}} = 300\,\text{mV}$, $I_{\text{stab}} = 200\,\text{pA}$, $V_{\text{mod}} = 1\,\text{mV}_{\text{rms}}$. $12 \times 12$ spectra were taken over an area of $300\,\text{nm} \times 300\,\text{nm}$. The subbands $E_{\text{0}}$ and $E_{\text{1}}$ are marked at $-60\,\text{mV}$ and $110\,\text{mV}$, respectively.\label{fig:dIdV1}}
\end{figure}
\begin{figure*}\capstart
\includegraphics{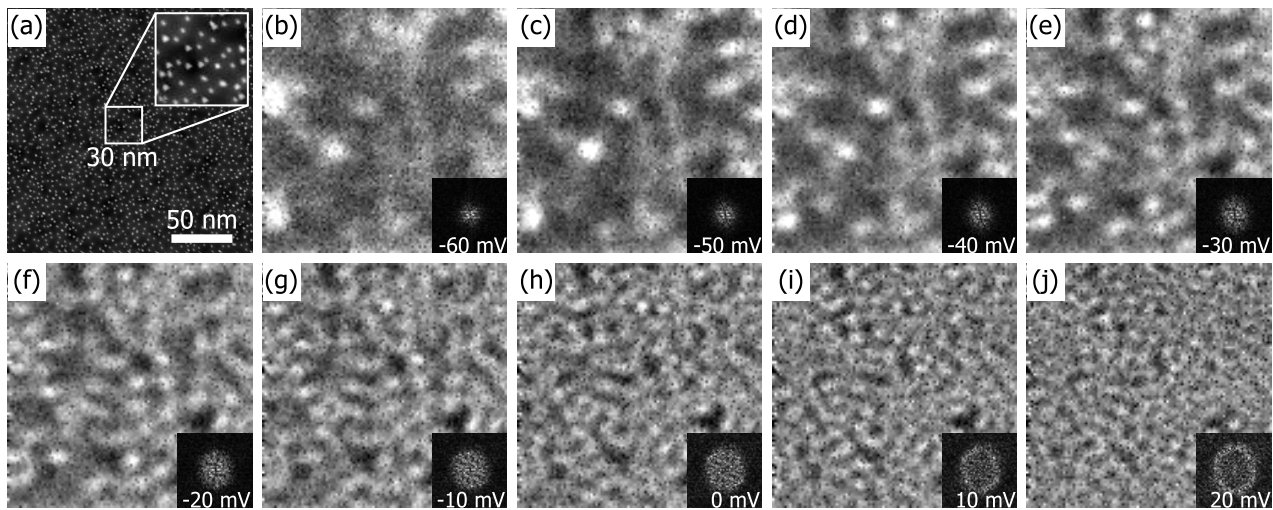}
\caption{ (a) STM constant-current image of InSb(110) covered with 1.5 \% of a monolayer Cs; Cs adsorbates appear as white dots; the size of the inset is indicated by the white rectangle; $200\,\text{nm} \times 200\,\text{nm}$, $V = 300\,\text{mV}$, $I = 100\,\text{pA}$; (b)--(j) $dI/dV(x,y)$ images  of the same surface area as shown in (a) recorded at $V=-60\,\text{mV}$ to $V=20\,\text{mV}$ as marked in the insets; $V_{\text{stab}} = 300\,\text{mV}$, $I_{\text{stab}} = 100\,\text{pA}$, $V_{\text{mod}} =1\,\text{mV}_{\text{rms}}$, pixel resolution: $2\,\text{nm}$; insets show Fourier transformations of the real-space images displaying the electron wave vectors contributing to the $LDOS(x,y)$.\label{Wellen}}
\end{figure*}
Figure~\ref{fig:dIdV1} (lower curve) shows the $dI/dV(V)$ spectrum of Cs covered InSb(110) at $B=0\,\text{T}$, which is averaged from 144 curves recorded on a regular grid covering an area of $(300\,\text{nm})^2$. The first and second subband can be identified as steps in the $dI/dV$ signal, which due to averaging represents the density of states (DOS). The applied sample voltage $V_{\text{s}}$ corresponds to the electron energy $E$ relative to the Fermi energy $E_{\text{F}}$. Obviously, only the first subband is occupied by electrons, since the second subband is located above $E_{\text{F}}$ ($V_{\text{s}} =0\,\text{mV}$). Assuming reasonably that the band bending induced by Cs is complete at $1.5\,\%$ coverage,\cite{NbInAs} i.e. $V_{\text{bb}} = 290\,\text{meV}$,\cite{PhysRevB.63.155315} the subband energies $E_i$ relative to $E_{\text{CBM}}$ are deduced to be $E_{\text{0}}=230\,\text{meV}$ and $E_{\text{1}}=400\,\text{meV}$. Using Eq.~(\ref{eq:Merkt}) with $m^*_0 = 0.0135 \times m_{0}$\cite{Vurgaftman} ($m_{0}$: free electron mass), the effective electron mass in the first subband increases from $0.022 \times m_{0}$ at the subband edge ($E_0$) towards $0.029 \times m_{0}$ at the Fermi level ($E_{\text{F}}$). The effective mass at the onset of the second subband ($E_1$) would be $0.042\times m_{0}$. Using further Eqs.~(\ref{eq:Ns}) and (\ref{eq:DE}), the electron concentration of the 2DES can be deduced to be $N_{\text{s}}=6.5 \times 10^{15}\,\text{m}^{-2}$. Notice, that the $dI/dV$ curve of Fig.~\ref{fig:dIdV1} (lower curve) does not show any states of a tip induced quantum dot,\cite{Tipinduced} which has been achieved by careful preparation of a tip with adaptive work function to the sample surface.

\begin{figure}\capstart
\includegraphics{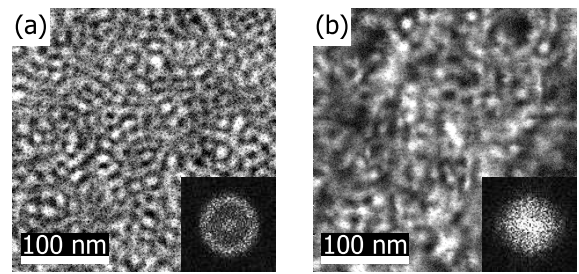}
\caption{(a), (b) $dI/dV(x,y)$ images recorded at the Fermi energy ($V_{\text{s}} = 0\,\text{mV}$) and at $B=0\,\text{T}$ (a) and $B=7\,\text{T}$ (b); $300\,\text{nm} \times 300\,\text{nm}$, $V_{\text{stab}} = 300\,\text{mV}$, $I_{\text{stab}} = 200\,\text{pA}$, $V_{\text{mod}} =1\,\text{mV}_{\text{rms}}$, pixel resolution: $1\,\text{nm}$; insets show Fourier transformations of real space data.\label{nullvolt}}
\end{figure}

The $dI/dV(x,y)$ ($LDOS$) images probed at energies within the first subband are shown in Fig.~\ref{Wellen}(b)--(j). One observes the evolution of standing wave patterns exhibiting decreasing wave length with increasing energy. These patterns are very similar to the ones observed within a 2DES induced by Fe on $n$-type InAs(110).\cite{PhysRevLett.89.136806} Notice that the small black dots appearing in all $dI/dV$ images at the same position are the Cs atoms, which appear dark due to the larger distance of the tip with respect to the InSb surface. The insets show the Fast Fourier Transformations (FFTs) of the real-space images displaying the electron wave vectors $\underline{k}$ contributing to the real-space pattern of the $LDOS$ with $\underline{k}=\underline{0}/\text{nm}$ being in the center. In the ideal case of a 2DES with negligible potential disorder, the FFT would show a ring, growing in diameter with the non-parabolic energy dispersion relation $E(k)$. The contribution of larger $k$ values with increasing energy can indeed be deduced from the growing disc in the FFTs at low energy, which develops into a ring structure at about $V_{\text{s}}=0\,\text{mV}$ still increasing in diameter with energy. The contrast of the ring can be improved by recording larger images with a higher spatial resolution as shown in Fig.~\ref{nullvolt}(a) for the wave pattern at $E_{\text{F}}$.
A clear ring structure appears in the FFT image, which is difficult to observe in Fig.~\ref{Wellen}(h). The transition from a disk-like appearance towards a ring-structure with remaining intensity in the center of the ring has been observed previously for a disordered 2DES.\cite{PhysRevLett.89.136806} There, it has been reproduced by Hartree calculations taking into account the potential disorder of the 2DES, produced by the charged dopants. For strongly disordered systems, one observes only the disk increasing in diameter with energy.\cite{PhysRevB.68.041402} Thus, the FFTs, not being a perfect ring structures, indicate the wave function mixing by the spatially fluctuating electric potential due to charged dopants. In accordance with expectations, the wave function mixing gets reduced with increasing energy.

\begin{figure}\capstart
\includegraphics{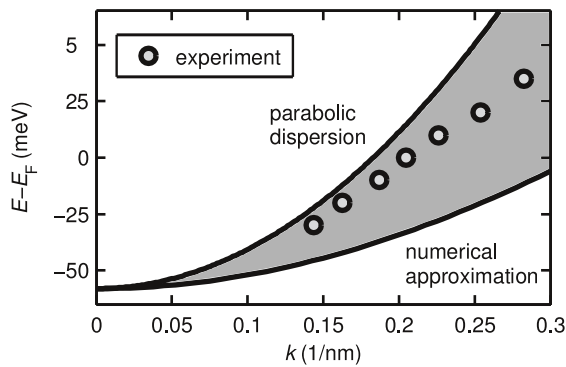}
\caption{The electron energy $E$ relative to the Fermi energy $E_{\text{F}}$ is plotted against the $k$ value parallel to the 2DES layer. The dots represent the dominant $k$ values extracted from Fig.~\ref{Wellen} (e)--(j). The upper line shows a parabolic dispersion with the estimated  effective mass of $m^{*}=0.022 \times m_{0}$ at the subband edge. The lower curve is a numerical solution of Eq.~(\protect{\ref{eq:MomEffMass}}) and Eq.~(\protect{\ref{eq:Merkt}}). The deviation from the parabolic dispersion marks the non-parabolicity of the 2D conduction band dispersion of InSb, the deviation from the numerical approximation is caused by the influence of higher order terms in $\textbf{k}\cdot\textbf{p}$ theory.\label{fig:dispersion}}
\end{figure}

Figure~\ref{fig:dispersion} shows the dominant $k$ values taken from the maxima in radial line scans of the FFT images in Fig.~\ref{Wellen} and \ref{nullvolt}. They are displayed as dots in comparison with a theoretical upper and lower limit of the expected InSb dispersion. The upper limit is a parabolic dispersion resulting from the effective mass $m^*=0.022 \times m_0$ at the onset of the 2DES $E_0$. It neglects the non-parabolicity within the 2DES. The lower curve is obtained by solving Eq.~(\ref{eq:MomEffMass}) and Eq.~(\ref{eq:Merkt}) numerically, which overestimates the non-parabolicity at the high energies of the 2D electrons by neglecting higher order terms of $\textbf{k}\cdot\textbf{p}$ theory.\cite{PhysRevB.35.2460,AndradaeSilva1994} A full numerical treatment of the $\textbf{k}\cdot\textbf{p}$ Kane model \cite{Winkler.PhysRevB.48.8918} is beyond the scope of this paper. Nevertheless, the experimental data points are found in between the two limits evidencing that the wave patterns of Fig.~\ref{Wellen} are indeed caused by the electrons of the 2DES.

\begin{figure}\capstart
\includegraphics{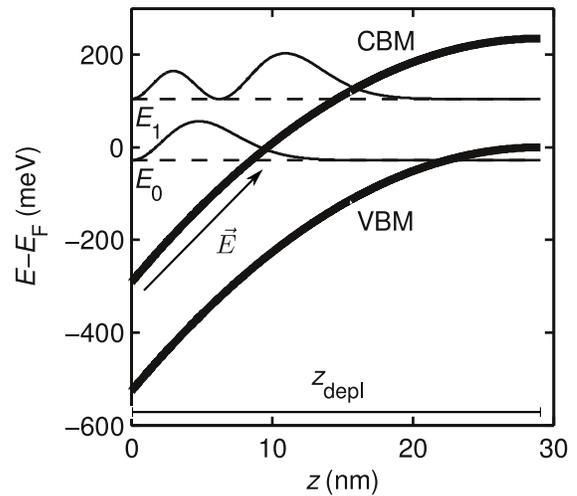}
\caption{Calculated band bending for the 2DES: conduction band minimum (CBM), valence band maximum (VBM) and subband energies $E_{0}$, $E_{1}$ are plotted as a function of distance $z$ from the surface. The assumed acceptor concentration and electron concentration of the 2DES are $N_{\text{A}} = 1 \times 10^{24}\,\text{m}^{-3}$ and $N_{\text{s}}=6.5 \times 10^{15}\,\text{m}^{-2}$, respectively, in accordance with experimental data. The electron distribution curves $|\Psi_i(z)|^2$ resulting from a triangular well approximation are drawn for each subband.\cite{RevModPhys.54.437} The depletion length $z_{\text{depl}} = 29\,\text{nm}$ and the electric field visible within the first subband $|\vec{E}|=3.1 \times 10^{7} \,\text{V/m}$ are additionally marked.\label{BandBending}}
\end{figure}

Interestingly, the wave pattern intensity is rather continuously distributed over the area of Fig.~\ref{Wellen} and \ref{nullvolt}(a). This can be most clearly seen in Fig.~\ref{nullvolt}(a) and is in contradiction to expectations from single point scattering, where a reduction of intensity $I$ with distance $r$ from the scatterer according to $I(r) \propto r^{-1}$ is expected.\cite{Kanisawa} One reason might be that this discrepancy is influenced by multiple scattering paths to the standing wave pattern, which are known to be considerably more important in 2D than in 3D.\cite{Anderson} Another reason might be the large number of scatterers within the image area of Fig.~\ref{Wellen}. This number can be estimated with the help of the depth of the 2DES as displayed in Fig.~\ref{BandBending} ($9\,\text{nm}$), the acceptor concentration ($10^{24}\,\text{m}^{-3}$) and the image size ($(200\,\text{nm})^2$). This results in 360 acceptors scattering the electron waves within Fig.~\ref{Wellen}.

\begin{figure}\capstart
\includegraphics{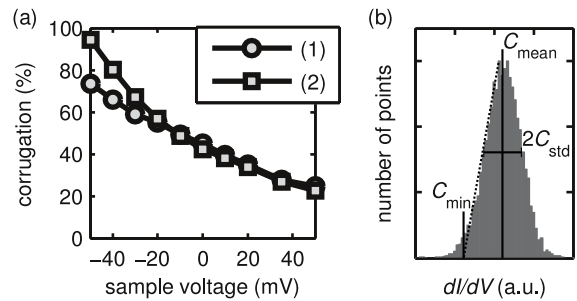}
\caption{(a) Corrugations $K$ of the electron wave patterns within the first subband; (1) $K:=(C_{\text{mean}}-C_{\text{min}}) / C_{\text{mean}}$, (2) $K:=(2 \times C_{\text{std}}) / C_{\text{mean}}$ with $C_{\text{std}}$, $C_{\text{mean}}$ and $C_{\text{min}}$ being standard deviation, mean and minimum in the intensity distribution of the $dI/dV$ images. (b) Histogram of Fig.~\ref{Wellen}(h) indicating the derived values. $C_{\text{min}}$ marks the $dI/dV$ offset from a linear fit of the left slope (dotted line) of the histogram which amounts to cutting off the lowest $2\,\%$ of $dI/dV$ values. $C_{\text{mean}}$ and $C_{\text{std}}$ are calculated numerically.\label{fig:Korrugation}}
\end{figure}

The corrugation $K$ of the wave patterns, which is caused by the strength of the scattering and the phase coherence length, is displayed in Fig.~\ref{fig:Korrugation}(a). It is calculated by two different methods using the histograms of $dI/dV$ values within an image as shown in Fig.~\ref{fig:Korrugation}(b). The first method determines the difference between the mean value $C_{\text{mean}}$ and the minimum value $C_{\text{min}}$ and divides it by $C_{\text{mean}}$: $K=(C_{\text{mean}}-C_{\text{min}})/C_{\text{mean}}$. The second method instead uses twice the standard deviation of the histogram $C_{\text{std}}$ and divides it by $C_{\text{mean}}$: $K=2 \times C_{\text{std}}/C_{\text{mean}}$. Both approaches give similar results except at low energy as visible in Fig. \ref{fig:Korrugation}(a). In previous STS studies, method (1) has been used revealing that a 2DES ($K=60\,\%$) is prone to a much larger $K$-value than a three-dimensional system ($K=3\,\%$), even if the potential disorder is quite similar.\cite{Morgenstern2003121} Moreover, a drop of $K$ from $90\,\%$ to $50\,\%$ has been found in a strongly disordered 2D system at the percolation transition of strongly localized states.\cite{PhysRevB.68.041402}
Our $K$-values are compatible with the previous values obtained on 2DESs, but, other than in the previous publications, $K$ becomes continuously smaller with increasing energy. This can be explained straightforwardly by the increase of the scattering length with energy. However, the difference with respect to the previous data is not completely clear.

Next, we discuss the STS data obtained in a perpendicular magnetic field of $B=7\,\text{T}$. The $dI/dV$ image recorded at the same sample position as Fig.~\ref{nullvolt}(a) using exactly the same tunneling parameters is shown in Fig.~\ref{nullvolt}(b).  The pattern looks much more disordered and a dominant wave length is not visible anymore as evidenced by the FFT in the inset. In a perpendicular magnetic field, Landau quantization and spin splitting is expected for 2D electrons. In the effective mass approximation neglecting Rashba spin splitting, the Landau and spin levels are given by
\begin{eqnarray}
E^{n}_{i,\pm}=E_{i} + \hbar\omega_{c} (n+\frac{1}{2}) \pm \frac{1}{2} g^{*} \mu_{\text{B}} B
\label{eq:Landau}
\end{eqnarray}
with subband index $i$, Landau level index $n=0,1,2,\ldots$, spin index $\pm$, Bohr magneton $\mu_{\text{B}}=e\hbar/2m_{0}$, elementary charge $e$, cyclotron frequency
\begin{eqnarray}
\omega_c = \frac{eB}{m^{*}},
\label{eq:omegac}
\end{eqnarray}
and effective Land\'e g-factor $g^{*}$. The spin splitting given by the third term in Eq.~(\ref{eq:Landau}) is also energy dependent within a non-parabolic conduction band, i.e. $g^*(E)$. An approximation for  $g^*(E)$ is given by the relation:\cite{Kanskaya}
\begin{eqnarray}
\frac{g^{*}(E)}{g^{*}_{0}} &=&  \frac{m^{*}_{0}}{m^{*}(E)}.
\label{eq:Kanskaya}
\end{eqnarray}
For InSb, one can use $g^{*}_{0}=-51$ being the effective g-factor at the conduction band edge.\cite{Vurgaftman}

\begin{figure}\capstart
\includegraphics[width=80mm]{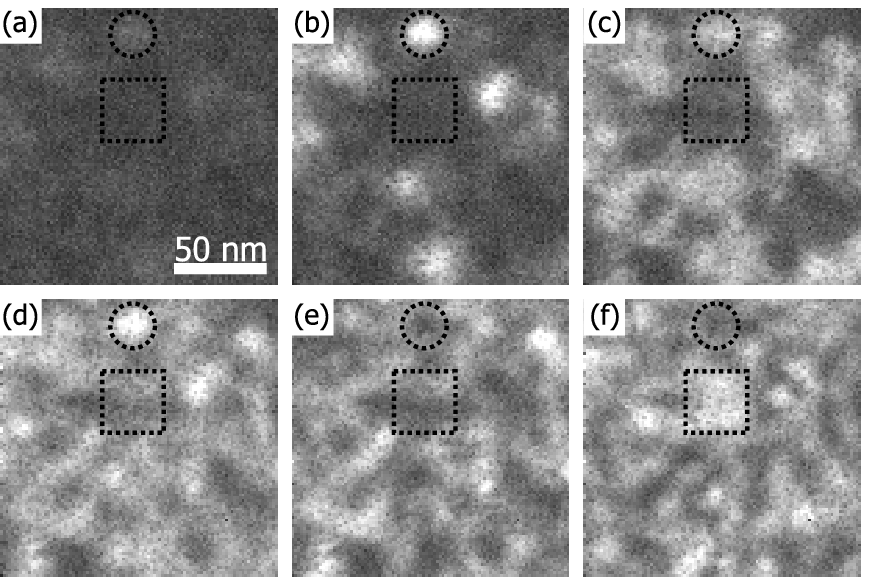}
\includegraphics{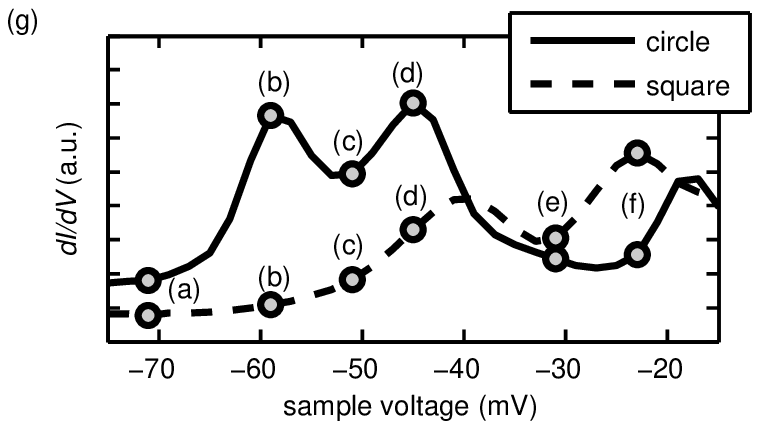}
\caption{
(a)--(f) $dI/dV$ images with $1.5\,\text{nm}$ resolution at $B=7\,\text{T}$, $150\,\text{nm} \times 150\,\text{nm}$, $V_{\text{stab}} = 300\,\text{mV}$, $I_{\text{stab}} = 200\,\text{pA}$, $V_{\text{mod}} = 1\,\text{mV}_{\text{rms}}$; the corresponding sample voltages are marked in the local  $dI/dV$ curves shown in (g); (g) local $dI/dV$ curves spatially averaged over the two small areas marked in the images (a)--(f).\label{Perkolation}}
\end{figure}
\begin{figure}\capstart
\includegraphics[width=80mm]{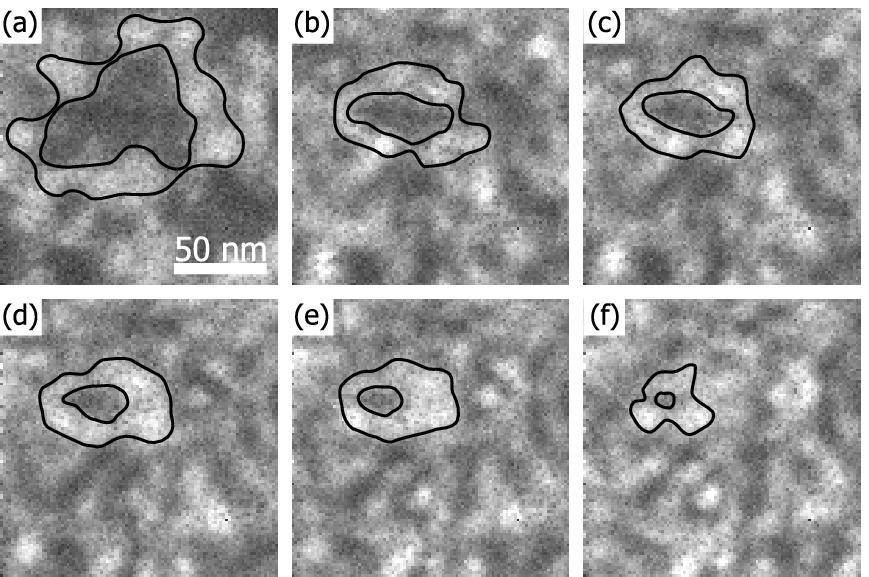}
\includegraphics{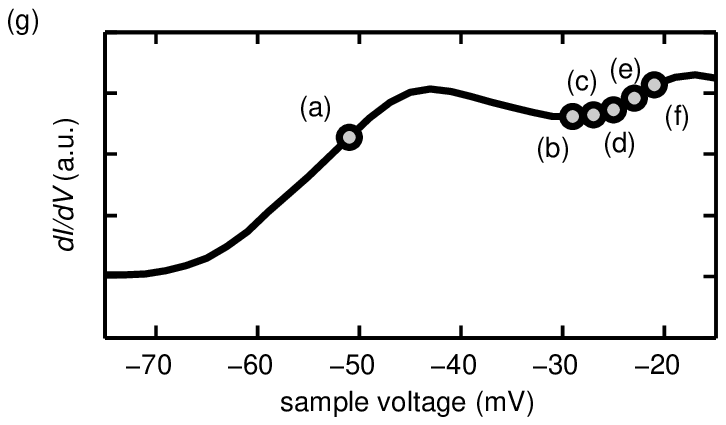}
\caption{(a)--(f) $dI/dV$ images recorded at $B=7\,\text{T}$, $V_{\text{stab}} = 300\,\text{mV}$, $I_{\text{stab}} = 200\,\text{pA}$, $V_{\text{mod}} = 1\,\text{mV}_{\text{rms}}$,  with different sample voltages as marked in (g), $150\,\text{nm} \times 150\,\text{nm}$; the intensity marked by black lines highlights a drift state moving uphill with increasing energy; (g) spatially averaged $dI/dV$ spectrum originating from the $10^4$ curves covering the whole image area of (a)--(f). \label{PerkolationZulaufen}}
\end{figure}
Within an electric potential landscape, the Landau states experience an additional electric field perpendicular to the $B$-field, which leads to drift states, i.e. states covering closed equipotential lines of the disorder.\cite{Prange,Ando2} At the lowest possible energy, the first Landau level starts with states at potential pits representing localized electrons. With increasing energy, the equipotential lines representing the states encompass an increasing area until they percolate at the critical energy close to the center of the Landau level. There, the extended critical state of the Quantum Hall transition appears. At even higher energy, the states localize again by covering equipotential lines circulating around potential maxima. This behavior, taking place in each Landau and spin level, has recently been observed  directly by STS at $T=0.3\,\text{K}$.\cite{hashimoto:256802} A similar percolation of states in $B$-field has also been observed on graphite surfaces.\cite{Nimii}

For our system, the transition is shown in Figs.~\ref{Perkolation} and \ref{PerkolationZulaufen} concentrating on the lowest Landau level. The images (a)--(f) in Fig.~\ref{Perkolation} display $dI/dV$ images at $B=7\,\text{T}$ at increasing sample voltage. The circle drawn in each image marks a potential pit, since it shows $dI/dV$ intensity already at the lowest energies. The square, in contrast, marks an area, where the potential is relatively high, i.e. the $dI/dV$ intensity appears only at rather high energy.  The curves taken at these positions are shown in Fig.~\ref{Perkolation}(g) and indeed exhibit a pair of peaks shifted by approximately $20\,\text{meV}$ with respect to each other. One can identify the doublet as the first spin split Landau level, which would result in $|g^{*}|=32$ for the spin splitting and $m^{*}=0.020\times m_{0}$ for the distance to the next Landau level observed within the pit. This is in excellent agreement with the values $|g^{*}|=31$ and $m^{*}=0.022\times m_{0}$ calculated for the subband edge $E_0$ from first-order $\textbf{k}\cdot \textbf{p}$-theory (Eqs.~\ref{eq:Merkt} and (\ref{eq:Kanskaya})). By comparing the two curves of Fig.~\ref{Perkolation}(g), we deduce that our 2DES exhibits a disorder potential with a potential fluctuation of about $20\,\text{meV}$ and, by regarding the distance of square and circle in Fig.~\ref{Perkolation}(a)--(e), we deduce a spatial length scale of the potential corrugation of about $30\,\text{nm}$.

Since the potential fluctuation is larger than the spin splitting, it is difficult to observe the appearance and disappearance of extended states. However, a first extended state is visible in Fig.~\ref{Perkolation}(c), where the states developing from the potential pits touch and form connecting paths from one side of the image to the other. An upward movement of a drift state onto a potential hill is highlighted in Fig.~\ref{PerkolationZulaufen}. The black lines roughly mark the $dI/dV$ intensity of a closed structure, which at low energy surrounds an area of about $(80\,\text{nm})^2$. With increasing energy, the dark inner area shrinks continuously condensing towards a small spot in (f). The overall increase of the spatially averaged $dI/dV$ signal in (g) can be attributed to the drift states of the next Landau level, which start to develop in the lower potential areas. The visibility is disturbed by the overlap with states from other levels. Therefore, Fig.~\ref{PerkolationZulaufen}(a) is taken from the lower spin level and Figs.~\ref{PerkolationZulaufen}(b)--(f) are taken from the higher spin level leading to reduced overlap with other states in both cases. Indeed, the energetic overlap of states from different Landau and spin levels at different potential energy completely removes the spin splitting from the spatially averaged $dI/dV$ curve, i.e. the DOS, and weakens the visibility of the Landau level splitting significantly, as shown in Fig.~\ref{PerkolationZulaufen}(g).

Finally, we discuss the observation of the Rashba effect within our 2DES. As described in Eq.~(\ref{eq:Landau}), constant values of $m^{*}$ and $g^{*}$  would lead to equally spaced Landau and spin levels at constant $B$-field. Each level is broadened by the disorder potential, which is relatively large in our system because of the high acceptor concentration prohibiting the observation of spin splitting in the spatially averaged density of states. For non-parabolic conduction bands, the Landau level splitting $\hbar\omega_{c}(E)$ and the spin splitting $|g^{*}(E)| \mu_{\text{B}} B$, both, decrease with energy. This fact, however, cannot explain the spatially averaged spectrum shown in Fig.~\ref{fig:dIdV1} (upper curve), which exhibits a beating pattern of the Landau level intensity within the first subband. The proposed explanation for the beating is Rashba spin splitting,\cite{BychkovRashba} which gives a spin splitting for confined 2D electrons moving perpendicular to an electric field even at $B=0\,\text{T}$. The electric field can be externally applied or is given by an asymmetric confinement potential. The Rashba spin splitting increases with wave number $k$, leading to two dispersion curves with different effective mass. Including this effect into the effective mass Hamiltonian results in Landau levels given by\cite{BychkovRashba}
\begin{eqnarray}
E^{n,\sigma}_{i}&=&E_{i} + \hbar\omega_{c} \left( n + \sigma \left( \delta^2 + \gamma^2 n  \right)^{1/2} \right), \label{eq:Rashba}\\
\gamma &=& \alpha \left( 2 m^{*} /\hbar^{3} \omega_{c} \right)^{1/2}, \\
\delta &=&  \frac{1}{2} \left( 1 - \frac{m^{*} g^{*}}{2 m_{0}} \right),
\end{eqnarray}
with $n = 0,1,2,\ldots$ being the level index, $\sigma = 1$ for $n = 0$, and $\sigma = \pm 1$ for $n = 1,2,3,\ldots$ being the spin index. The parameter $\alpha$ is called the Rashba parameter. Here it encodes the interaction of the conduction band with the spin-orbit split valence band and can be deduced from $\textbf{k}\cdot \textbf{p}$-theory.\cite{KaneInSb,AndradaeSilva1994} One can easily show that Eq.~(\ref{eq:Rashba}) gives the same energy levels as Eq.~(\ref{eq:Landau}) for $\alpha = 0$. It is obvious from Eq.~(\ref{eq:Rashba}), that the g-factor induced spin splitting given by $\delta$ is the same for all $n$, while the $\alpha$ dependent splitting encoded in $\gamma$ increases with the level index $n$. This eventually leads to a mixing of spin states from different $n$ in the density of states inducing a beating.

\begin{figure*}\capstart
\includegraphics{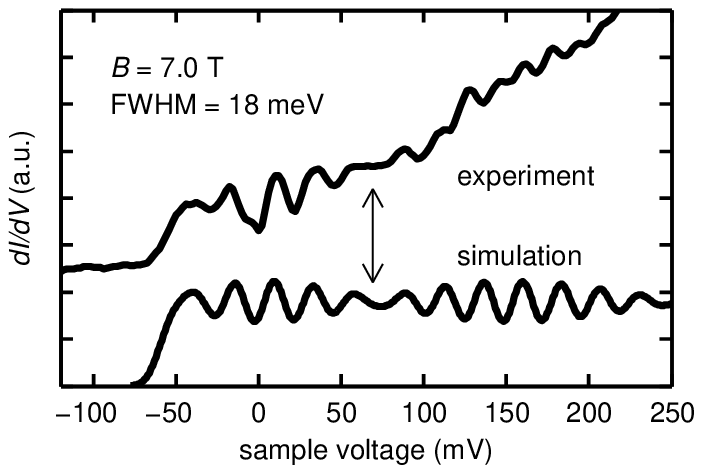}
\includegraphics{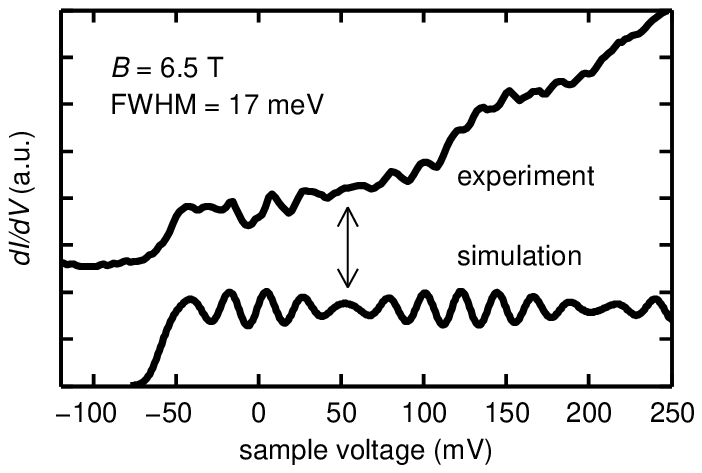}
\includegraphics{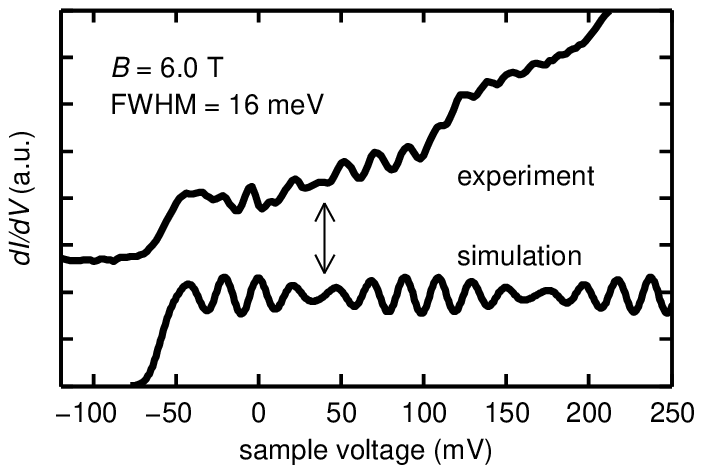}
\includegraphics{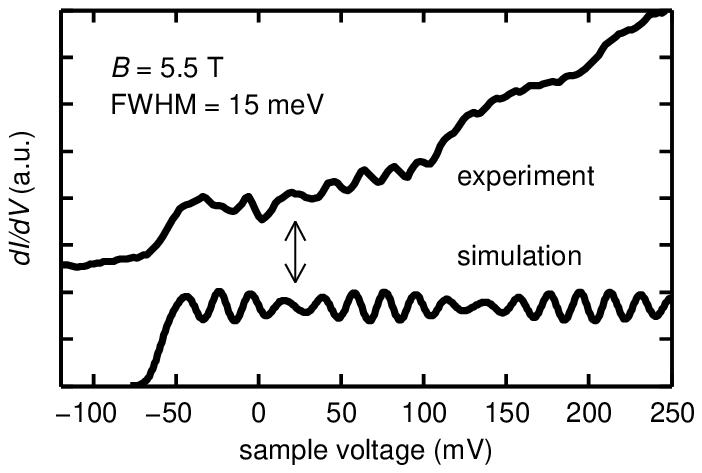}
\includegraphics{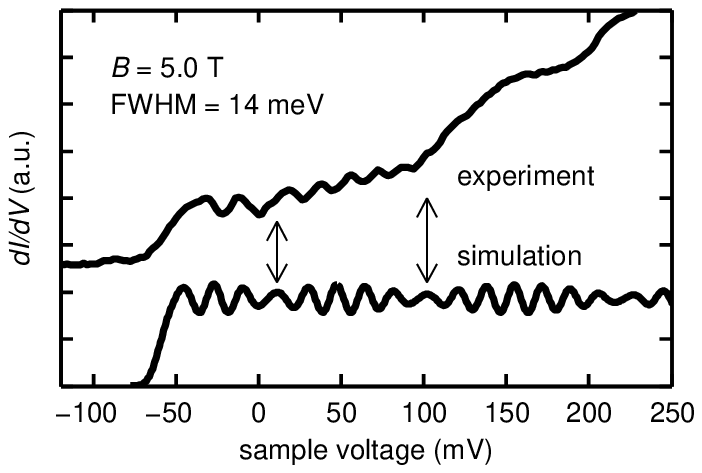}
\includegraphics{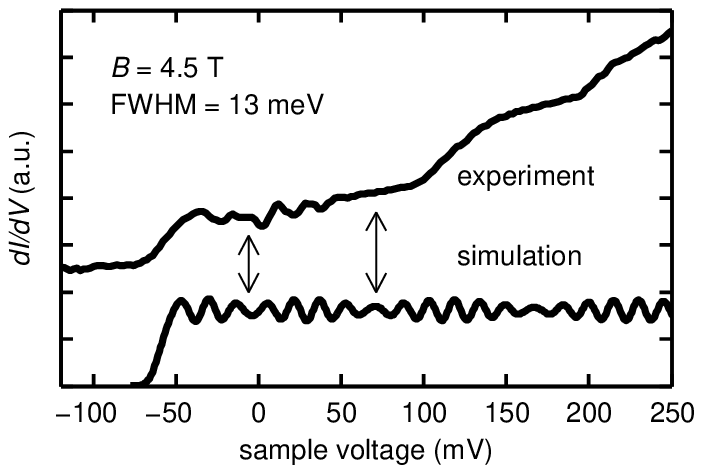}
\caption{Spatially averaged $dI/dV$ (DOS) spectra at different $B$-fields as marked (experiment) in comparison with simulations using Landau and spin energies from Eq.~(\ref{eq:Rashba}) with $m^{*}=0.035 \times m_{0}$, $\alpha=7 \times 10^{-11}\,\text{eV}\,\text{m}$, $g^{*}=-21$ and a Gaussian broadening with full width at half maximum (FWHM) as indicated for each $B$-field; the beating minima are marked by arrows.\label{fig:Rashba}}
\end{figure*}
The Rashba parameter $\alpha$ for the 2D conduction band electrons of InSb can be estimated in an eight band Kane model\cite{KaneInSb} with a confining potential $V(z)$,\cite{AndradaeSilva1994} thereby describing non-parabolicity as the interaction of the spin split conduction band with the nearest three spin split valence bands, of which one is separated by the spin-orbit splitting energy $\Delta=0.8\,\text{eV}$. This results in an energy and $z$ dependent Rashba parameter:\cite{AndradaeSilva1994}
\begin{eqnarray}
\alpha(z,E) &=& \frac{P^2}{2}\frac{d}{dz}\left(\frac{1}{E-V(z)+E_{g}}\right. \nonumber\\
&&\left.-\frac{1}{E-V(z)+E_{g}+\Delta}\right), \label{eq:Silva}\\
P^2&=&\frac{\hbar^2}{m^{*}_{0}}\frac{E_{g} \left(E_{g} + \Delta \right) }{3 E_{g}+2 \Delta}.
\end{eqnarray}

To calculate the spin splitting via Eq.~(\ref{eq:Rashba}), $\alpha(z,E)$ has to be convolved with the wave function of the lowest subband in $z$-direction $\Psi_0(z)$. In order to obtain $\Psi_0(z)$, the confining potential is calculated as shown in Fig.~\ref{BandBending}. For this calculation, the maximum cesium induced band bending $V_{\text{bb}}=290\,\text{mV}$ has been used to solve the Poisson equation. The screening charge is given by the bulk acceptor concentration $N_{\text{A}}$, which is negatively charged by electrons from the adsorbates, and the estimated inversion-layer electron concentration $N_{\text{s}}=6.5 \times 10^{15}\,\text{m}^{-2}$.\cite{RevModPhys.54.437} We find, that $N_{\text{s}}$ has a negligible influence on the resulting confinement due to the dominating 2D acceptor concentration $N_{\text{A}} \times z_{\text{depl}} = 3.0 \times 10^{16}\,\text{m}^{-2}$. Using the triangular well approximation with an infinite barrier at $z=0\,\text{nm}$\cite{RevModPhys.54.437} and the effective mass given by Eq.~(\ref{eq:Merkt}), two subbands result within the confinement potential exhibiting energies in good agreement with the experimental values shown in Fig.~\ref{fig:dIdV1}. More precisely, the calculated energies $E_{0}$ and $E_{1}$  are by about 10 \% larger than the experimental ones, which is not surprising, because the simple model neglects penetration of the electrons into the vacuum and into the valence band, which both would lower their energy.   A more accurate calculation would require solving the Poisson equation and a multi-band Kane Hamiltonian self-consistently,\cite{marques1982theory} but this is beyond the scope of this paper. Calculating the band bending for the whole width of possible acceptor concentrations $N_{\text{A}} = 1\text{--}2 \times 10^{24}\,\text{m}^{-3}$ reveals electric fields within the 2DES of $|\vec{E}|=3.1\text{--}4.1 \times 10^{7} \,\text{V/m}$. $\vec{E}$ as well as the square of the wavefunctions $|\Psi_{i}(z)|^2$ resulting from the triangular well approximation are drawn in Fig.~\ref{BandBending} highlighting the nearly constant value of $\vec{E}$ across $|\Psi_0(z)|^2$.

Finally, the effective Rashba parameter $\alpha$ for the first subband is determined convolving the result of Eq.~(\ref{eq:Silva}) with $\Psi_0(z)$ numerically:
\begin{eqnarray}
\alpha=\langle \Psi_{0}(z)| \alpha(z,E_{0}) | \Psi_{0}(z) \rangle_{z} .
\end{eqnarray}
This results in a Rashba parameter  $\alpha=9\text{--}11 \times 10^{-11}\,\text{eV}\,\text{m}$ being higher than the values observed in InAs inversion-layers or heterostructures by transport measurements ($3\text{--}4 \times 10^{-11}\,\text{eV}\,\text{m}$).\cite{PhysRevB.61.15588} It should be noted that the calculated Rashba parameter is an upper estimate, since the ignored effect of barrier penetration of the electronic waves decreases the effective Rashba parameter. Furthermore, $\alpha$ is only the lowest order of an inversion asymmetry induced spin splitting and it is known that higher orders lead to a reduced effect.\cite{Winkler.PhysRevB.48.8918} Indeed, the observed beating patterns displayed for different $B$-fields in Fig.~\ref{fig:Rashba} can be reproduced nicely by using Eq.~(\ref{eq:Rashba}) with a slightly reduced Rashba parameter of $\alpha=7 \times 10^{-11}\,\text{eV}\,\text{m}$. For the sake of simplicity, we assumed a constant effective mass and $g^*$-factor taken as the average value within the first subband ($m^{*}=0.035\times m_{0}$, $g^*=-21$) and we fit a Gaussian broadening to each spin and Landau level accounting for the potential disorder in order to obtain the calculated curves in Fig.~\ref{fig:Rashba}. Note that the fit parameter $\text{FWHM} = 13\text{--}18\,\text{meV}$ given in Fig.~\ref{fig:Rashba} nicely agrees with the potential disorder deduced independently from Fig.~\ref{Perkolation}.

Although we could not reproduce all details of the measured spectra, it is obvious that the minima of the beating (nodes), marked by arrows in Fig.~\ref{fig:Rashba}, are in excellent agreement with the experiment. The calculation also reproduces the increase of beat frequency with decreasing $B$-field. The decreased FWHM found for lower $B$ reflects the fact that the lateral extension $d$ of the drift states scales according to $d\propto 1/\sqrt{B}$.\cite{Prange,Ando2,Morgenstern2} Thus, drift states become more insensitive to the steepest parts of the fluctuating disorder potential at lower $B$. We checked carefully, that the chosen FWHM does not influence the position of the beating nodes.

The Rashba spin splitting is, of course, also present at $B=0\,\text{T}$ and would split an approximated parabolic energy dispersion $E(k)$ into two branches separated by $\pm \alpha k$,\cite{BychkovRashba} if higher order spin splitting terms are neglected. This leads to a spin splitting of about $30\,\text{meV}$ at $E_{\text{F}}$ and two $k_{\text{F}}$ values of about $k_- =2.7\times 10^8\,\text{m}^{-1}$ and $k_+ =2.1\times 10^8\,\text{m}^{-1}$. However, these two $k_{\text{F}}$-values are not visible in the FFT of Fig.~\ref{nullvolt} in accordance with theory.\cite{Pettersen} Slight changes within the complex $LDOS$ pattern might appear, if multiple scattering is involved,\cite{Walls} but these changes could only be pinpointed by detailed comparison with $LDOS$ calculations including the details of the disorder potential.\cite{Morgenstern2003121} A more detailed theoretical consideration of the consequences of disorder on the Rashba-split $LDOS$ based on previous calculations\cite{Walls,PhysRevB.71.241312} might be an interesting base for such experiments. Notice that the width of the ring in the FFT of Fig.~\ref{nullvolt}(a) is as large as the difference of $k_+$ and $k_-$ at $E_{\text{F}}$ showing that the Rashba effect and the disorder within our sample are of the same strength. We also checked, if a direct observation of zero field splitting is possible within the density of states as has been recently proposed by Ref.~[\onlinecite{PhysRevLett.89.096805}], but without success.

\section{Conclusion}
In summary, we have successfully prepared a 2DES close to the surface of a highly doped $p$-type InSb crystal exhibiting one occupied subband only. Using scanning tunneling spectroscopy, the evolution of standing waves in the 2DES has been imaged being dominated by wave numbers explained by a non-parabolic dispersion relation, but exhibiting strong wave function mixing and large corrugation due to disorder. In magnetic field, Landau and spin levels are observed locally, but washed out in the density of states due to the disorder with amplitudes of about $20\,\text{meV}$. Percolating drift states are observed within the potential landscape. The density of states shows an irregular Landau level pattern dominated by beating which could be attributed to Rashba spin splitting caused by the asymmetry of the confining potential. The deduced Rashba parameter of $\alpha=7 \times 10^{-11}\,\text{eV}\,\text{m}$ is relatively large and very close to the value of $\alpha=9\text{--}11 \times 10^{-11}\,\text{eV}\,\text{m}$ estimated by $\textbf{k}\cdot \textbf{p}$-theory. This shows that Rashba parameters can also be determined by a local probe, which provides spatial resolution down to the atomic scale. This opportunity might trigger novel types of experiments within the near future.

\begin{acknowledgments}
We want to thank Mike Pezzotta and Katsushi Hashimoto for their valuable help during the experiments.
\end{acknowledgments}

%\bibliography{pInSb}
%Merlin.mbs v4.21 2009-07-09.
%

\end{document}